\documentstyle[12pt]{article}

\input amssym.def
\input amssym.tex
\newcommand{\starh}{*_{\hbar}}
\newcommand{\pten}{\Pi} 
\newcommand{\inv}{^{-1}}
\newcommand{\cala}{{\cal A}}
\newcommand{\calh}{{\cal H}}
\newcommand{\call}{{\cal L}}
\newcommand{\calp}{{\cal P}}
\newcommand{\calt}{{\cal T}}
\newcommand{\bolda}{{\bf A}}
\newcommand{\boldm}{{\bf M}}
\newcommand{\boldp}{{\bf P}}
\newcommand{\pf}{\noindent{\bf Proof.}\ }
\newcommand{\del}{\partial}
\newcommand{\qed}{\begin{flushright} $\Box$\ \ \ \ \ \
                  \end{flushright}}
\newcommand{\defequal}{\stackrel{\mbox {\tiny {def}}}{=}}
\newcommand{\pcomm}[2]{\{[#1,#2]\}}
\newcommand{\dpcomm}[2]{\{\{[#1,#2]\}\}}
\newcommand{\pbr}[2]{\{#1,#2\}}
\newcommand{\dpbr}[2]{\{\{#1,#2\}\}}
\newcommand{\cinf}{C^{\infty}}
\newcommand{\reals}{\Bbb R}
\newcommand{\frakg}{{\frak g}}
\newcommand{\calf}{{\cal F}}
\newtheorem{thm}{Theorem}[section]
\newtheorem{prop}[thm]{Proposition}

\title{{\bf Geometry of the transport equation in multicomponent WKB
approximations}}
\author{Claudio Emmrich and Alan Weinstein \thanks{The research of both authors
was partially supported by DOE Contract DE-FG03-93ER25177.  We would like to
thank
Hans Duistermaat, Mikhail Karasev, Robert Littlejohn, and Jim Morehead for
helpful
discussions.}\\Department of Mathematics
\\University of California\\
Berkeley, CA 94720 USA\\
{\small \tt emmrich@math.berkeley.edu~~~alanw@math.berkeley.edu }}
\date{December 14, 1994}
\begin{document}
\maketitle

\begin{abstract}
Although the WKB approximation for multicomponent systems has been
intensively studied in the literature, its geometric and global aspects
are much less well understood than  in the scalar case. In this paper
we give a completely geometric derivation of the transport equation,
without using local sections and without assuming complete diagonalizability
of the matrix valued principal symbol, or triviality of its eigenbundles.
The term (called ``no-name term'' in some previous literature) appearing in the
transport equation in addition to the
covariant derivative with respect to a natural projected connection
will be a tensor, independent of the choice of any sections.
We give a geometric interpretation of this tensor, involving the
contraction of the curvature of the eigenbundle and an analog of the
second fundamental
form with the Poisson tensor in phase space. In the non-degenerate
case this term may be rewritten in an even simpler
geometric form. Finally, we discuss  obstructions to the existence
of WKB states and give a  geometric description of the
quantization condition for WKB states for a non-degenerate
eigenvalue-function.
\end{abstract}

\section{Introduction}
In its original analytic form, the so-called WKB method for obtaining
asymptotic
eigenfunctions for linear partial differential operators involves writing a
trial
approximate eigenfunction for an operator $H$ in the form
$\psi(x)=e^{iS(x)/\hbar}a(x)$.  Expanding $H\psi-E\psi$ in powers of
$\hbar$ leads first to a nonlinear first order partial differential equation
(the
eikonal, or Hamilton-Jacobi equation) for the phase function $S$ and then to a
linear
homogeneous first order partial differential equation for the amplitude $a$
(the
transport equation).

A geometric version
of the WKB method was developed by Maslov \cite{ma:perturbations} and
H\"{o}rmander \cite{ho:fourier}, in which the phase
function is represented by a lagrangian submanifold $L$ in classical phase
space, and
the amplitude by a half-density $\alpha$ on $L$.  This geometric approach makes
it
possible to extend the WKB method to cover in a natural way the so-called
caustic
points, which are inevitable in bound-state problems, and which lead to
singularities
in the analytic approach.  We refer to \cite{ba-we:lectures} \cite{du:fourier}
\cite{gu-st:geometric} for extensive treatments of this ``geometric WKB''
theory.

Much of the work described in the preceding paragraphs was carried out
originally for equations in a single unknown (complex-valued)
function.  Since many interesting equations in mathematical physics
involve several functions (or even sections of nontrivial vector
bundles), it has been of interest to extend the WKB method to such
multicomponent equations.  Much progress has been made in this
direction, both in pure mathematics and in mathematical physics
(see for instance
\cite{co:dirac}\cite{de:propagation}\cite{ka:global}
\cite{ka-ma:operators}\cite{ma-fe:semiclassical}).
For the physical approach, we refer especially to
\cite{li-fl:geometric}, which contains extensive references to earlier work,
and
which was the starting point for our own study.  The cited authors have
extended
the analytic version of the WKB method to the
case where the symbol matrix of the differential equation at hand has
an eigenvalue of constant multiplicity.  Their results are general enough to
cover the local theory in the presence of caustics, but a completely
geometric description has not been achieved, in particular for the
transport equation.

The aim of this paper is to present a geometric formulation, with a
coordinate-free, gauge-invariant derivation, of the transport equation
for multicomponent systems in terms of a linear connection on the
eigenvector bundle of the principal symbol matrix of a system of
linear differential operators.  We compare our results with those of
\cite{li-fl:geometric}, showing in particular that the ``no-name term'' in
their formulation of the transport equation can be interpreted in
terms of the curvature of the eigenvector bundle and its complement.  More
precisely, the connection involves a contraction of the curvature with
the Poisson tensor on phase space.  This same contraction appears in K\"ahler
geometry
\cite{ko:differential}, where it is known as ``mean curvature'' and appears to
involve more
structure; we note here that it really depends only on the Poisson structure
associated to
the K\"{a}hler form and therefore call it  ``Poisson
curvature''.  Its importance in the context of symplectic geometry
is only now becoming apparent.  (See \cite{ya:yang}.)

The ultimate goals of our study go beyond the scope of this paper.  One is to
clarify the semiclassical quantization conditions in the multicomponent
setting.  Another is to deal with the extremely important ``level-crossing''
problem in which the eigenvalues of the symbol matrix have variable
multiplicity.  We hope that our geometric methods will facilitate work on these
difficult problems (see \cite{br-du:normal} and \cite{li-fl:phase} for recent
contributions), even though we do not attack them here.

\section{Projection matrices over the Moyal algebra}
\label{sec-projection}

Our basic strategy will be the same as that used in much of the previous
work--to
begin by breaking off from the given operator a piece corresponding to the
eigenspaces in question, and then to consider the reduced system, whose
principal
symbol is a multiple of the identity matrix by a scalar function.  In this way,
we reduce the problem as far as possible to the scalar case.  Interesting
geometry arises from the fact that the natural domain of this
reduced system is a vector bundle over phase space which is locally
``twisted''.
When the phase space has non-trivial topology, this bundle can also be globally
nontrivial.  (Even when the original phase space is topologically simple, we
may
have to remove points at which the multiplicity of the eigenvalue increases,
 leaving behind a topologically complicated space.)

Our treatment will differ from that in \cite{li-fl:geometric} in
that we do not attempt to put the operator in question into block
diagonal form (which requires a choice of eigenvectors which depends
smoothly on points in phase space); instead we follow
\cite{co:dirac}  \cite{ka:global} \cite{ka-ma:operators} by putting the
emphasis on projections onto the
eigenspaces.  These latter objects are completely canonical.

As in \cite{li-fl:geometric}, we will use the calculus in which operators are
represented by matrices whose entries are formal power series in a small
parameter
(we use $\hbar$ instead of their $\epsilon$) whose coefficients are $\cinf$
functions
on classical phase space.  The operation of these functions on phase
space will be by the Weyl ordering, so the appropriate multiplication
of the matrix entries is by the
Moyal product.  We will need to make explicit use of only the initial part of
the
development of this product,
\begin{equation}
\label{eqn-moyal}
a \starh b = ab + (i\hbar/2) \pbr{a}{b} + O(\hbar),
\end{equation}
so that although the Moyal product itself applies only to the phase space
$\reals^{2n}$, the results in this section of our paper will be applicable
whenever we are dealing with a phase space carrying a Poisson bracket
satisfying
the usual axioms
\cite{va:lectures}. Note that the commutator bracket $[a,b]_*=a\starh b -
b\starh a$ is
asymptotic to
$i\hbar\pbr{a}{b}$ as $\hbar \rightarrow 0$.

Before going further, let us fix some terminology and notation.  We denote by
$\bolda$ the algebra of functions on phase space, with the usual pointwise
multiplication.  $\bolda[[\hbar]]$ denotes the algebra of formal power series
in
$\hbar$ with coefficients in $\bolda$, with multiplication given by the Moyal
product.  $\boldm_N$ denotes the algebra of $N\times N$ matrices with
coefficients in $\bolda$. Its elements can also be thought of as matrix-valued
functions on phase space, with the multiplication given by pointwise matrix
multiplication.  We will often write $\boldm$ for $\boldm_N$ when it is not
important to specify the dimension of the matrices.  Finally,
$\boldm_N[[\hbar]]$ (or $\boldm[[\hbar]]$ for short) denotes
the space of formal power series with coefficients in $\boldm_N$, with
multiplication given by thinking of its elements as matrices with
entries in
$\bolda[[\hbar]]$.  This multiplication is also given by a formula
like (\ref{eqn-moyal}) above, where the first term is ordinary matrix
multiplication, and the Poisson bracket of matrix-valued functions is
defined by $\pbr{a}{b}_{ij}=\sum _{k}\pbr{a_{ik}}{b_{kj}}.$

The hamiltonian $H$ which we will consider will be an element of
$\boldm [[\hbar]]$.  Given any such element $A=A_0(x)+\hbar A_1(x) +
\cdots$, we call the matrix-valued function
$A_0(x)$ the {\bf principal symbol} of $A$.  A scalar function
$\lambda(x)$ will be called a {\bf regular eigenvalue function} for
$H$ if $\lambda(x)$ is an eigenvalue for $H_0(x)$ with multiplicity
independent of $x$, and if the null space and range of
$H_0(x)-\lambda(x)I$ are complementary subspaces for each value of
$x$.  (The latter condition is satisfied automatically if the values
of $H_0$ are hermitian matrices.)  There is a well-defined projection
matrix $\pi_0(x)$ onto the $\lambda(x)$-eigenspace along the range of
$H_0(x)-\lambda(x)I,$ which depends smoothly on $x$.  The images of
the $\pi_0$ (the family of eigenspaces of the $H_0(x)$) form a vector
bundle over phase space which we denote by $E_\lambda$ and call the
$\lambda$-{\bf eigenbundle} of $H_0$.  $E_\lambda^\perp$ will denote
the family of null spaces of the $\pi_0(x)$.  It is a complementary
bundle to $E_\lambda$; it really is an orthogonal complement when the
$H_0(x)$ are hermitian operators and the $\pi_0(x)$ are consequently
orthogonal projections.

For spectral theory, we will need a $\pi\in \boldm [[\hbar]]$ which is a
projection
in the sense that $\pi \starh \pi =\pi $ and whose principal symbol is
$\pi _{0}$.  Such a projection always exists and can even be chosen to
commute with $H$, according to the following Proposition.

\begin{prop}
\label{prop-projections}
Let $\lambda $ be a regular eigenvalue function for $H$, $\pi _{0}$
the corresponding projection onto the eigenbundle $E_{\lambda }$ along
$E_{\lambda }^\perp$.  Then there is a unique projection  $\pi \in
\boldm [[\hbar]]$
whose principal symbol is $\pi _{0}$ and which commutes with $H$.
\end{prop}

\pf
We first show that we can modify $\pi_0$ by adding higher order terms
such that it becomes a projection with respect to the *-product.
 To see that, we use an
induction argument and suppose that we have chosen elements $\pi
_{1},\ldots \pi _{k}$ of $\boldm$ such that $\pi ^{(k)}=\pi
_{0}+\hbar \pi _{1}+\cdots +\hbar^{k}\pi _{k}\in \boldm[[\hbar]]$ is a
projection through order $k$, i.e. $(\pi^{(k)})^{2}-\pi^{(k)} =
\hbar^{k+1}a_{k+1}+O(\hbar^{k+2})$.  We wish to choose $\pi _{k+1}$ so
that $\pi ^{(k+1)}=\pi ^{(k)}+\pi _{k+1}$ is a projection through
order $k+1$.  This requires us to solve the equation $$a_{k+1} +
\pi_0\pi_{k+1}+\pi_{k+1}\pi_0 = \pi_{k+1},$$ or equivalently
$$\pi_0\pi_{k+1}-\pi_{k+1}(1-\pi_0) = -a_{k+1}.$$ Now the operator
$p\mapsto\pi_0 p- p(1-\pi_0)=\pi_0 p\pi_0 -(1-\pi_0 )p(1-\pi_0 )$ maps all
matrix valued functions to
those which are block diagonal with respect to the splitting
determined by $\pi _{0}$ (i.e. those commuting with $\pi _{0}$), and
annihilates the matrices which are strictly off-diagonal with
respect to this splitting.  So it suffices to show that $a_{k+1}$
commutes with $\pi _{0}$.  But $a_{k+1}$ is the principal symbol of
$\hbar^{-(k+1)}((\pi^{(k)})^{2}-\pi^{(k)})$, which commutes with
$\pi^{(k)}$, hence $a_{k+1}$ commutes with the principal symbol $\pi
_{0}$ of $\pi^{(k)}$.  Thus we can always choose a suitable (non-unique)
 $\pi _{k+1}$, and, by induction,  there is always a projection with
 principal symbol $\pi_0$.

To prove our proposition, we use a second induction argument.
Suppose that we have chosen a projection $\pi^{(k)}$ so that
$[H,\pi^{(k)}]_*=O(\hbar^{(k+1)}).$  (To start, we take an arbitrary
projection with principal symbol  $\pi^{0}$  , as constructed above.)  Then
there is a unique
$F\in \boldm $ such that
$$[H,\pi^{(k)}]_*=\hbar^{k+1}F+O(\hbar^{k+2}).$$  We will choose the next
approximation to have the form (exponentials are with respect to the Moyal
product)
$$\pi^{(k+1)}=e^{\hbar^{k+1}A}\starh\pi^{(k)}\starh e^{-\hbar^{k+1}A}$$ which
is
automatically a projection for any $A\in \boldm $.  Expanding the
exponentials gives
$$\pi^{(k+1)}=\pi^{(k)}+ \hbar^{k+1}[\pi_0,A]_*+O(\hbar^{k+2}),$$ so
$$[H,\pi^{(k+1)}]_*=[H,\pi^{(k)}]_*+
\hbar^{k+1}[H,[\pi_0,A]_*]_*+O(\hbar^{k+2}),$$ which equals $\hbar^{k+1}(F+
[H_0,[\pi_0,A]])+O(\hbar^{k+2}).$  So we must choose $A$ as a solution of the
equation $F+ [H_0,[\pi_0,A]]=0.$  This is possible as long as $F$ is
off-diagonal with respect to the block decomposition given by
$\pi_0$, i.e. if
$\pi_0 F \pi_0$ and $ (1-\pi_0) F (1-\pi_0) $ vanish.\footnote{To see this, one
needs
 only to use the invertibility of
$Q-\lambda I$, where $Q=(I-\pi_0)H_0(I-\pi_0)$ is the compression of $H_0$ to
the
range of $H_0-\lambda I$.}

But these matrix functions are
nothing but the principal symbols of the operators
$\pi^{(k)}\starh [H,\pi^{(k)}]_* \starh\pi^{(k)}$ and
$(1-\pi^{(k)})\starh[H,\pi^{(k)}]_*\starh (1-\pi^{(k)}),$ which vanish just
because
$\pi^{(k)}$ is a projection.

The uniqueness of $\pi $ is proven by a similar stepwise argument,
using the two requirements that it should
be a projection and commute with $H$.

\qed

We remark that, if we have several regular eigenvalue functions
$\lambda _{\mu },$ then the corresponding projections $\pi _{\mu }\in
\boldm [[\hbar]]$ will all commute with one another.  In particular,
if all the eigenvalues of $H_{0}$ are regular, we have a complete
decomposition into ``polarization sectors.''  We would like to stress,
though, that this complete decomposition is not as essential for the
study of a single eigenvalue function as it may appear to be from some
formulas in \cite{co:dirac} and \cite{ka:global}.

The projection $\pi _{0}$ in $\boldm $ has been chosen so that $(H_{0}-\lambda
I)\pi _{0}$ and $\pi _{0}(H_{0}-\lambda
I)$ both vanish.  As a result, for the projection $\pi $  in $\boldm [[\hbar]]$
constructed in Proposition \ref{prop-projections}, the elements
$(H-\lambda  I)\starh\pi $ and $\pi\starh (H-\lambda I)$
of $\boldm [[\hbar]]$ are both of order $\hbar$, as is $\pi \starh(H-\lambda
I)\starh \pi $.  Since $\pi \starh(H-\lambda I)\starh\pi $ commutes with $\pi $
and is annihilated by left or right multiplication by $(I-\pi )$, its
leading order term has the same properties with respect to $\pi _{0}$.

Let us compute it: the coefficient of $\hbar$ in $\pi\starh (H-\lambda
I)\starh\pi$ is
$$ \pi _{0}H_{1}\pi _{0} + \pi _{1}(H_{0}-\lambda I)\pi _{0}+ \pi
_{0}(H_{0}-\lambda I)\pi _{1}
  +\frac{i}{2}(\pbr{\pi _{0}}{H_{0}-\lambda I}\pi_0 +\pbr{\pi
_{0}(H_{0}-\lambda )
}{\pi _{0}}).$$
All but two of the terms vanish, and we can always add extra factors
of $\pi _{0}$ on the outside, so we conclude:
$$ \pi \starh (H-\lambda I) \starh \pi =\hbar\pi _{0}(H_{1}+\frac{i}{2}
\pbr{\pi
_{0}}{H_{0}-\lambda I})\pi _{0}+O(\hbar^{2}). $$

Notice in particular that $\pi _{1}$ has disappeared entirely from
this expression.

\section{WKB Approximation}
\label{sec-WKB}

We will seek a WKB eigenfunction for $H$ which is in the image of the
projection $\pi $ found above.  Specifically, we choose a lagrangian
submanifold $L$ in phase space on which the eigenvalue function
$\lambda $ has the constant value $E$, and a ``principal symbol'' $u$ on
$L$, which is a section of the tensor product of the half-densities on
$L$ with the vector space ${\Bbb C}^{n}$.
The Maslov procedure associates to this data an $\hbar$-dependent wave
function $\psi $ for which $u $ is called the principal symbol.  For
instance, if $L$ has the form $p=dS(q)$ for a phase function $S$ on
configuration space, we can take $q$ as a coordinate on $L$ and write $u$
in the form $a(q)\sqrt{|dq|}$, where $a$ is a vector-valued function.
The associated wave function is then
$\psi = e^{\frac{i}{\hbar}S(q)}a(q)\sqrt{|dq|}$.

What is important is
not so much the specific form of the WKB ansatz but the fact that,
when we apply an operator $A$ to such a $\psi $, the result is again
associated to $L$, with the principal symbol $A_{0}u$.  In the special
case that $A_{0}u=0$ and $A_0$
is a scalar function multiple $a_0I$ of the identity matrix, $A\psi$ is of
order $\hbar$, and $\hbar\inv A\psi$ has principal symbol $A_1 u
-i\call_{X_{a_0}} u$, where the second term is $-i$ times the Lie derivative of
$u$ by the hamiltonian vector field of $a_0$.  (This vector field is tangent
to $L$ because $L$ is a lagrangian submanifold on which the function $a_0$
vanishes.)   We also note that the Moyal product on  $\boldm[[\hbar]]$ is
consistent with its operation on wavefunctions:
$(A\starh B)\psi = A(B\psi)$.

In particular, by
applying the projection
$\pi
$ to wavefunctions associated with
$L$, we obtain (all the) wavefunctions which are in the image of $\pi
$ and thus candidates for the approximate eigenfunctions which we are
seeking.

Suppose then that $\pi \psi =\psi $.  Since $H$ commutes with $\pi $,
we have $H\psi =H\pi \psi =\pi H\psi =\pi H\pi \psi. $  Therefore, by
the main result of the previous section,
$H\psi =(\pi \lambda I  \pi + \hbar\pi _{0}(H_{1}+\frac{1}{2} i\pbr{\pi
_{0}}{H_{0}-\lambda I})\pi _{0})\psi +O(\hbar^{2}).$  This means that, as far
as
its action on $\psi$ is concerned, the operator $H$ can be replaced by one
whose
principal symbol is the scalar multiple $\lambda I$ of the identity,
and we can apply the standard analysis in this special case.

 Now let
$E$ be a candidate for an eigenvalue for
$H$.  The order 0 part of $(H-EI)\psi $ is then $(\lambda -E)\psi $, which we
can
kill by choosing $\psi $ to be associated with a lagrangian submanifold
contained in the level surface for the value $E$ of the eigenvalue
function $\lambda $, which now plays the role of a scalar hamiltonian
for our purposes.

The transport equation for the symbol $u$ of $\psi $ is the
requirement that the principal symbol of
$\hbar\inv(H-EI)\psi $ be zero.  This principal symbol is
$$\pi _{0}(H_{1}+\frac{1}{2} i\pbr{\pi _{0}}{H_{0}-\lambda I})\pi
_{0})u - i\call_{X_{\lambda}}u)$$ plus the principal symbol of
$\hbar\inv\pi \starh(\lambda -E)\psi .$

Modulo $O(\hbar^2),$ $$\pi \starh(\lambda -E)\psi = (\pi_0+
\hbar\pi_1)\starh(\lambda -E)\psi .$$  Since $(\lambda -E)\psi$ is already of
order $\hbar$, this reduces to  $(\pi_0\starh(\lambda -E))\psi ,$  in which the
coefficient of $\hbar$ is $\frac{i}{2}\pbr{\pi_0}{\lambda}\psi$.  After further
application (always permissible) of the projection $\pi_0$, this becomes zero,
since $\pi_0\pbr{\pi_0}{\lambda}\pi_0=0$.\footnote{The argument is as follows.
$\pbr {\pi_0}{\lambda}=\pbr {\pi_0^2}{\lambda}=\pbr {\pi_0}{\lambda}\pi_0+\pi_0
\pbr {\pi_0}{\lambda}$.  Multiplying on the left and right by $\pi_0$ gives
$\pi_0
\pbr {\pi_0}{\lambda}\pi_0 =2\pi_0
\pbr {\pi_0}{\lambda}\pi_0$, so $\pi_0 \pbr {\pi_0}{\lambda}\pi_0=0$. }

We can now write the transport equation for the symbol $u$, a half-density on
$L$ with values in the $\lambda$-eigenbundle:

\begin{equation}
\label{eqn-transport1}
 \pi_0\call_{X_{\lambda}}u +((-1/2)
\pi_0\pbr{\pi _{0}}{H_{0}-\lambda I}\pi_{0}+i\pi_0 H_1\pi_0)u=0.
\end{equation}

\section{Geometric interpretation}
\label{sec-geometric}

In this section, we will give a geometric interpretation of the terms in the
transport equation in the language of connections on vector bundles and their
curvature.

If we write the symbol $u$ as $ a \otimes \nu$ for a complex-valued
half-density
$\nu$ on $L$ and a section  $a$ in the $\lambda$-eigenbundle over $L$,
 the first term of the transport equation (\ref{eqn-transport1}) becomes:
\[ \pi_0\call_{X_{\lambda}}u =  a \otimes  \call_{X_{\lambda}}\nu +
D_{X_{\lambda}} a \otimes \nu ,
\]
where $D$ is the covariant differentiation on  sections of the
$\lambda$-eigenbundle defined by
$D \zeta = \pi_0 d \zeta $ for an arbitrary section $\zeta$.  D is the
covariant
differential associated with the connection on $E_{\lambda}$ naturally
associated with the trivial connection on the trivial ${\Bbb C}^N$ bundle over
 phase space (having $d$ as its covariant differential) and the projection
$\pi_0$
from the trivial bundle to the eigenbundle.  It was observed by Simon
\cite{si:holonomy} that such projected connections, which are standard in
differential geometry, especially the geometry of submanifolds (see for example
\cite{bi-cr:geometry}), are just the ones whose holonomy in certain
situations of physical interest is popularly called Berry's phase, after
\cite{be:quantal}.  Thus, corresponding expressions in the transport equation
are
named ``Berry'' terms in \cite{li-fl:geometric}.

We turn next to the matrix-valued function in the second term on the left
hand side of (\ref{eqn-transport1}).  It corresponds to the ``no-name'' terms
in
\cite{li-fl:geometric}, but we will denote it by $\Lambda_C$ and call it the
{\bf curvature term}, for reasons which will become clear shortly.

The curvature term may be rewritten as follows:
\begin{eqnarray}
 \Lambda_{C} \defequal  \pi_0 \pbr{\pi _0}{H_0-\lambda I} \pi_{0} &=&
  \pi_0\pbr{\pi_0}{\lambda (\pi_0 -I)} \pi_0 + \pi_0 \pbr{\pi_0}{ H_0 - \lambda
\pi_0}\pi_0
\nonumber \\
& = & \lambda  \pi_0\pbr{\pi_0}{\pi_0} \pi_0 +
\pi_0 \pbr{\pi_0}{ H_0 - \lambda \pi_0} \pi_0 . \label{eq:no-name}
\end{eqnarray}
We remark that both terms on the right hand side of (\ref{eq:no-name})  behave
tensorially when we multiply
$H_0(x)$ (and at the same time its eigenvalue function $\lambda(x)$) by a
function $f(x)$.  For the first term this is completely obvious; for the second
term it follows  from
$(H_0 - \lambda \pi_0)\pi_0=0$.

To give a geometrical interpretation of these tensorial terms, we
compute the curvature $F$ of the projected connection $D$. If we consider $D$
as a
covariant exterior derivative, $F$ is the 2-form  with values in the
endomorphisms of $E_{\lambda}$ for which
$   D^2 \psi=F \psi  $ for an arbitrary section $\psi$ of $E_{\lambda}$.
Since $D^2 \psi=\pi_0  d (\pi_0 d (\pi_0\psi) ) = \pi_0  (d \pi_0) \wedge (d
\pi_0)  \pi_0 \psi ,$ we find that $F= \pi_0  (d \pi_0) \wedge (d \pi_0) \pi_0
$.
  Hence, we see that the first term in
(\ref{eq:no-name}) is simply $\lambda <\Pi, F>$, where $<\Pi,F>$ denotes the
contraction of the Poisson tensor with the curvature 2-form.

To describe the second tensorial term geometrically, we first introduce an
analog
of the second fundamental form for embedded submanifolds: It is a 1-form with
values in the vector-bundle homomorphisms from the $\lambda$-eigenbundle
$E_{\lambda} $ to the kernel $E_\lambda^{\perp} $ of $\pi_0$ defined  by
$ S \zeta = (I -\pi_0) d \zeta $
for an arbitrary section of $E_{\lambda}$. Since $$S\zeta=(I -\pi_0) d \zeta =
(I -\pi_0) d (\pi_0\zeta)=
(I -\pi_0)\pi_0 d\zeta + (I -\pi_0)(d\pi_0)\zeta$$ and $(I -\pi_0)\pi_0=0,$
we have
$S=(I -\pi_0) (d \pi_0) $ which indeed takes values in the vector-bundle
homomorphisms. It  measures the extent to which the ``trivial'' parallel
transport defined by $d$ tends to move a vector out of the
$\lambda$-eigenbundle
into its complement, i.e. the discrepancy between the connections $d$ and $D$
when applied to sections of $E_{\lambda}$.

Similarly, we can define a 1-form with values in the homomorphisms from
$E_\lambda^{\perp}$ to $E_{\lambda}$ by
\[ S^* \eta = - \pi_0 d\eta = - \pi_0 d( I-\pi_0) \eta \]
for a section $\eta$ of $E_\lambda^{\perp}$. As the notation suggests, $S$ and
$S^*$ with the  above choice of sign are adjoint to one another if $\pi_0$ is
an orthogonal
projection on a hermitian vector
 bundle (e.g., if $H_0$ is hermitian).

Using $S$ and $ S^*$, we can define a 2-form with values in the endomorphisms
of $E_{\lambda}$ as $S^* \wedge ((H_0- \lambda \pi_0)  S)$,
where $H_0 - \lambda \pi_0$  is  considered as an endomorphism
of $E_{\lambda}^{\perp}$ (where it is just the restriction of $H_0$). This
2-form
can be contracted with the Poisson tensor to yield the missing term in
the transport equation. Indeed:
\begin{eqnarray*} \pi_0 \pbr{\pi_0}{ H_0 - \lambda \pi_0} \pi_0 &=&
 -\pi_0 \pbr{I- \pi_0}{( H_0 - \lambda \pi_0) (I-\pi_0) } \pi_0 \\&=&
- < \Pi, \pi_0 d(I- \pi_0) \wedge ( H_0 - \lambda \pi_0) d (I - \pi_0)  \pi_0 >
\\&=& < \Pi, \pi_0 d(I- \pi_0) \wedge ( H_0 - \lambda \pi_0) (I - \pi_0) d
\pi_0 >.
\end{eqnarray*}

Thus the curvature term is a sum:

\begin{equation}
\label{eq-curvature}
\Lambda_{C} =\lambda <\Pi, F> - <\Pi, S^* \wedge (H_0- \lambda \pi_0)  S >
\end{equation}

In general, the curvature term will be an endomorphism of an $m$-dimen\-sional
vector bundle (represented with respect to a local basis by an $m
\times m$ matrix), where $m$ is the dimension
of the $\lambda$-eigenspace, and it is not possible to simplify
further the
terms in the transport equation. However, in the special case that the
eigenvalue
function is non-degenerate (i.e., the multiplicity is 1), we can simplify them
by
observing that $\Lambda_{C}$ is uniquely determined  by  its trace
(here, we can compute
the trace on the whole vector bundle, not just on the eigenbundle,
since both give the same result!)
 and using the
invariance of the trace under cyclic permutation of factors. (The trace
operation does not
act on the  form part, so we just have to remember signs when we cyclically
change the order of forms).  We find that  $\Lambda_{C} =
\Lambda_{C}^{(s)} \pi_0 $, where the scalar  $\Lambda_{C}^{(s)}$ is
given by
\begin{eqnarray*}
 \Lambda_{C}^{(s)} &\!\!=\!\!& < \Pi, \mbox{tr} \Bigl( \lambda \pi_0  (d \pi_0)
\wedge (d
\pi_0) - \pi_0 d(I- \pi_0) \wedge ( H_0 - \lambda \pi_0) d (I - \pi_0)  \pi_0
\Bigr)>
\\ &\!\!=\!\!&  < \Pi, \mbox{tr} \Bigl(\lambda \pi_0  (d \pi_0) \wedge (d
\pi_0)
- \pi_0 (d\pi_0) \wedge ( H_0 - \lambda \pi_0) (d \pi_0)  \Bigr) > \\
&\!\!=\!\!&  < \Pi, \mbox{tr} \Bigl( \lambda \pi_0  (d \pi_0) \wedge (d \pi_0)
- (d\pi_0) \wedge ( H_0 - \lambda \pi_0) (d \pi_0) \Bigr)  > \\
&\!\!=\!\!& < \Pi, \mbox{tr} ( H_0 \tilde{F}) >
\end{eqnarray*}
where $\tilde{F} = d \pi_0 \wedge d \pi_0 $.
(In the second term we first used the cyclicity to get rid of the projection at
the end,
then $d \pi_0 = \pi_0 d \pi_0 + (d \pi_0) \pi_0$.)

$\tilde{F}$ is simply the curvature of the new connection $\tilde{D}$
on the trivial bundle defined by
\[ \tilde{D} \xi = \pi_0 d (\pi_0 \xi) + (I - \pi_0) d ((I-\pi_0) \xi ) \]
for an arbitrary section $\xi$. This {\bf adapted connection} (see
\cite{bi-cr:geometry}) preserves both the subbundles $E_{\lambda } $
and $E_{\lambda }^{\perp}$, its restriction to $E_{\lambda  }$ is just
$D$; in particular $F$ is simply the $\lambda$-block of $\tilde{F}$.

In order to compare our expression with those given in the literature,
and in particular that in \cite{li-fl:geometric}, we assume
that our hamiltonian is hermitian so that $\pi_0$ is an
orthogonal projection onto the one-dimensional eigenbundle $E_{\lambda
}$, in which we choose
a normalized local section $\tau$. Then $\pi_0 = \tau \tau^{\dagger}$,
and a straightforward calculation\footnote{We have borrowed here from
some notes of Jim Morehead.} (using $ \tau^ {\dagger} d \tau = - d(\tau^
{\dagger} ) \tau $, which follows from $ \tau^ {\dagger} \tau =1$)
yields
\[ \tilde{F} = \tau ( d \tau^ {\dagger}  \wedge d \tau) \tau^ {\dagger}  + d
\tau \wedge d\tau^ {\dagger}
 - [d \tau \wedge d\tau^ {\dagger}  ,   \tau \tau^{\dagger} ]_+  \]
where $[,]_+$ denotes the anticommutator. If we multiply by $H_0$ and take the
trace,
the anticommutator term vanishes, and  we finally get
\[\Lambda_{C}^{(s)}  = \lambda \pbr{ \tau^ {\dagger} }{\tau} +
 \sum_{\alpha \beta} (H_0)_{\alpha \beta} \pbr{ \tau_\alpha}{\tau^ {\dagger}
_\beta}\]
which is exactly the result in  \cite{li-fl:geometric}.

\section{Existence of quasiclassical states}
In WKB theory for the scalar case, one seeks quasiclassical eigenstates as
suitable vector valued half-densities on  lagrangian submanifolds $L$ of phase
space.  In attempting to extend this theory to the multicomponent case, one
encounters three difficulties: the presence of the
curvature term
$\Lambda_C$;  the fact that, even  if the curvature term vanishes, the
quasiclassical states are required to be covariantly constant along
hamiltonian trajectories with respect to a connection which is generally not
flat; and finally the fact that the  holonomy of this connection, even when it
is flat,  takes values  not in
${\Bbb C}^*$ or
$U(1)$ but in  $GL(m)$ or $U(m)$ (the latter if the projection $\pi_0$ is
orthogonal), where $m$ is the multiplicity of $\lambda$.

The curvature term presents a  problem mainly in the case of a
degenerate eigenvalue function $\lambda$.  In the non-degenerate case it is
simply
a scalar multiple of $\pi_0$, and hence can be replaced by a scalar,
$\hbar$-dependent part of the scalar hamiltonian; this is obviously not
possible
in the degenerate case.
Even in the non-degenerate case, the presence of the curvature term means that
the scalar hamiltonian is $\hbar$-dependent even if the matrix valued symbol is
$\hbar$-independent, which leads to the necessity of admitting
$\hbar$-dependent
lagrangian submanifolds \cite{li-fl:geometric}

The non-flatness of the connection makes it impossible to impose a naive
analog of the Bohr-Sommerfeld quantization condition, since the parallel
transport around cycles depends on the cycles
themselves, not just on their homotopy classes.

Whereas the two first problems  might be avoidable by  a suitable modification
of the geometric description of a quasiclassical state  (admitting
$\hbar$-dependent
lagrangian submanifolds and symplectic structures, and possibly making use
of a suitable extended phase space), the third problem is a real obstruction to
the
existence of quasiclassical states in the case of a degenerate eigenvalue
function.
  If we admit $\hbar$-dependent lagrangian submanifolds $L(\hbar)$
 as in \cite{li-fl:geometric}, the transport
equation (\ref{eqn-transport1}) will be modified, but only by an additional
$U(1)$ phase.
Hence, if we write the symbol $u$ as $a \otimes \nu$ for a complex-valued
half-density $\nu$
and a section $a$ in the $\lambda$-eigenbundle over $L= L(0)$, then  the
transport equation
for the corresponding section
$[a]$ in the projective $\lambda$-eigenbundle will be independent of $\hbar$.
 Hence, we have to find a section in the
projective eigenbundle which satifies this transport equation.
  Due to the $U(m)$-holonomy,
such a section will not always exist, even if the eigenvalue function $\lambda$
is
integrable. If the flow on the corresponding invariant torus is only
quasiperiodic,
it can  come arbitrarily close to a given starting  point without  the
correspondingly
transported point in the projective eigenspace being close to its starting
value.

This argument shows that the integrability of the eigenvalue function $\lambda$
is not a
sufficiently strong condition
 for the existence of a global WKB state, and in order to find an analog for
the quantization
condition for scalar systems one has to formulate a suitable strong notion of
integrability
for the classical limits of multicomponent systems.

In spite of the problems just listed, quasiclassical states can be shown to
exist in certain
cases, the easiest one being  that where the underlying phase space
is only two-dimensional  \cite{ka:global} and $H$ is hermitian.  In this case,
$L$ is
one-dimensional, so problems with the non-flatness of the connection do not
arise.  A
suitable section $[a]$ in the projective eigenbundle always exists.  To
construct it, one
simply chooses a point
$p$ on $L$, computes the holonomy around a loop based at $p$, selects for
$[a](p)$ the ray corresponding to one of the eigenvalues of the holonomy (which
is always
diagonalizable as it is unitary for hemitian
$H$), and defines $[a]$ by the transport equation. Hence, in this case a
quasiclassical
state exists, and the only effect of the non-trivial connection and the
curvature term is an
additional scalar phase which modifies the Bohr-Sommerfeld condition and is of
the same
order as the Maslov correction.

 In \cite{ka:global} the existence of quasiclassical states is shown for
certain other  examples
as well, where
the obstructions above are avoided by assuming either that either phase space
is
two-dimensional, that the fibers of the eigenbundle are only (complex)
one-dimensional, or
that the curvature term vanishes and that there is an ``adiabatic
connection''\footnote{We think that ``adiabatic constraint'' would be a better
translation of the original Russian in this instance.} -- i.e.,  a subbundle of
the
eigenbundle which is invariant and flat under the projected connection.

In the non-degenerate case, where the $\lambda$-eigenbundle is simply
a line bundle, there {\em does} exist a general method
for deriving a quantization condition.  Such a method is given in
\cite{li-fl:geometric}
for this special case,
    using local sections, diagonalization, and ``non-canonical
coordinates''. (In a somewhat different context, a similar result has
been obtained in \cite{ka:global}). In purely geometric terms their method  for
a phase space $T^*M$ with its canonical symplectic structure $\omega$ can be
described in
the following way.

 If we include the factor $e^{i S/\hbar}$ in the geometric description, the
quasiclassical states
are half-densities on a  submanifold of phase space  with values  in the tensor
product of the standard
trivial prequantum line bundle over a cotangent bundle and  the
$\lambda$-eigenbundle.
(We neglect the Maslov  correction for the moment).  In the non-degenerate case
this
bundle is again a line bundle,  and we can identify its curvature with a
two-form $F$ on
phase space. Since the connection on the prequantum bundle has curvature
$\frac{1}{\hbar}
\omega$,
 the curvature of the tensor product bundle is $\frac{1}{\hbar} \omega + F$.
Hence,
if we equip phase space with the  modified symplectic structure $\omega_\hbar =
\omega + \hbar F$,
then the curvature vanishes on the pullback of the line bundle above to any
submanifold $L_\hbar$ of $T^* M$ which is lagrangian with respect to
$\omega_\hbar$.
Hence, parallel sections exist at least locally on $L_\hbar$. In particular,
WKB states
correspond to $\hbar$-dependent lagrangian submanifolds contained in level sets
$\lambda_\hbar^{-1}(E)$ of the $\hbar$ dependent scalar hamiltonian function
$\lambda_\hbar$
obtained by including the curvature term.

Since the curvature 2-form vanishes on $L_\hbar$, we can formulate a
quantization condition
for cycles in the usual way (including the Maslov correction), which only
depends on the
homotopy class of the cycle. $L_\hbar$ will tend in the limit $\hbar
\rightarrow 0$ to a
submanifold $L_0$  which is lagrangian with respect to the unmodified
symplectic structure
$\omega$, and
 the Maslov correction can be computed from the
corresponding Maslov indices of $L_0$.
Thus, in the non-degenerate case it is possible to give a completely geometric
description
of quasiclassical states using globally defined objects.

The approach just described appears to apply only in the non-degenerate case.
Nevertheless, we expect that the purely  geometric derivation of the transport
equation can
serve  as a guideline to a formulation of a quantization condition in the
general case.

\end{document}